\documentclass[prl,twocolumn]{revtex4}
\usepackage{graphicx,amsmath,amssymb,amsxtra}
\usepackage{color}
\def\bd{\begin{displaymath}}
\def\be{\begin{equation}}
\def\ed{\end{displaymath}}
\def\ee{\end{equation}}
\def\bsub{\begin{subequations}}
\def\esub{\end{subequations}}

\newcommand{\Fig}[1]{Fig.~\ref{#1}}

\makeatletter
\newcommand*\dashline{\rotatebox[origin=c]{90}{$\dabar@\dabar@\dabar@$}}
\makeatother

\begin{document}

\title{Spin Correlations and Topological Entanglement Entropy in a Non-Abelian Spin-1 Spin Liquid}

\author{Julia Wildeboer and N. E. Bonesteel}

\affiliation{
Department of Physics and National High Magnetic Field Laboratory, 
Florida State University, Tallahassee, Florida 32310, USA}

\begin{abstract}
We analyze the properties of a non-Abelian spin-1 chiral spin liquid state proposed by Greiter and Thomale [PRL 102, 207203 (2009)] using variational Monte Carlo. In this state the bosonic $\nu = 1$ Moore-Read Pfaffian wave function is used to describe a gas of bosonic spin flips on a square lattice with one flux quantum per plaquette. For toroidal geometries there is a three-dimensional space of these states corresponding to the topological degeneracy of the bosonic Moore-Read state on the torus.  We show that spin correlations for different states in this space become indistinguishable for large system size. We also calculate the Renyi entanglement entropy for different system partitions to extract the topological entanglement entropy and provide evidence that the topological order of the lattice spin-liquid state is the same as that of the continuum Moore-Read state from which it is constructed.
\end{abstract}
\maketitle

{\em Introduction.}---The notion of quantum spin liquids, possible ground states of frustrated quantum antiferromagnets with no conventional long-range magnetic order, can be traced back to the original triangular lattice RVB state proposed by Anderson \cite{anderson1}.  Spin liquid ground states of gapped Hamiltonians in two dimensions are now understood to exhibit so-called topological order \cite{wen}, a type of order characterized by fractionalized excitations and nontrivial ground state degeneracies on topologically nontrivial surfaces.  Examples of theoretically established gapped spin liquids (and closely related dimer liquids) include the dimer liquid ground states on the triangular/kagome lattice \cite{MS, misguich}, an $SU(2)$-invariant $\mathbb{Z}_2$ quantum spin liquid on the kagome lattice \cite{seidel,wildeboer,wildeboer2}, and the Abelian chiral spin liquid (CSL) introduced by Kalmeyer and Laughlin \cite{KL1,KL2}. 

The Abelian CSL state is a spin-1/2 spin-liquid constructed using a quantum Hall type wave function for bosons that describes the amplitudes for spin-flips on a lattice and remains one of very few examples of a $2D$ spin liquid with fractional quantization.  In this Letter we investigate properties of a spin-1 non-Abelian CSL state proposed by Greiter and Thomale \cite{GT}. Both the Abelian and non-Abelian CSL states are derived from continuum wave functions, with the non-Abelian CSL emerging from a continuum Moore-Read state \cite{MR} known to have non-Abelian topological order with Ising anyon excitations. Entanglement properties on the cylinder studied in \cite{glasser} suggest these lattice CSL states harbor the same topological order as their continuum parents.   Here, we study the non-Abelian CSL 
for both planar and toroidal geometries and provide compelling evidence that it is indeed a quantum spin liquid with exponentially decaying spin correlations and that it possesses the topological order and associated modular $\mathcal{S}$-matrix of the continuum Moore-Read state. 

{\em Non-Abelian CSL state on planar geometry/torus.}---We begin by reviewing the spin-1 non-Abelian CSL state for planar geometry proposed by Greiter and Thomale \cite{GT}.  This state is constructed using the bosonic Moore-Read state \cite{MR} with filling fraction $\nu=1$ for which the droplet wave function in the symmetric gauge is, 
\begin{eqnarray}\label{wf_droplet_1}
\Psi[z_i] = {\rm Pf}\Big(\frac{1}{z_j - z_k}\Big) \prod_{i<j}^{N}(z_i - z_j) \prod_{i}^{N} e^{-|z_i|^2/4}.
\end{eqnarray}
We work in units with magnetic length equal to $1$ for which the square lattice formed by points with complex coordinates $z=\eta_{nm} = \sqrt{2\pi}(n+im)$ where $n$ and $m$ are integers has one flux quantum per plaquette.  If the bosons are restricted to this lattice, given the analytic structure of (\ref{wf_droplet_1}) each site can only have boson occupancies 0, 1, and 2, and, because the filling factor is $\nu=1$, there will be an average of one boson per site.

The spin-1 CSL state constructed using (\ref{wf_droplet_1}) is \cite{GT},
\begin{eqnarray}\label{wf_droplet_2}
|\Psi \rangle = \sum_{z_{1}, \ldots, z_{N}} \Psi[z_i] \prod_{i}^{N}G(z_i) \tilde{S}_{z_{1}}^{+} 
\ldots \tilde{S}_{z_{N}}^{+}|-1\rangle_{N},
\end{eqnarray}
where the $z_i$'s are summed over all lattice points $\eta_{nm}$.
Here $G(\eta_{nm}) = (-1)^{(n+1)(m+1)}$ is a gauge phase, and the operators $\tilde{S}_{z}^{+}$ are renormalized spin-flip operators, 
\begin{eqnarray}\label{operator_vacuum}
\tilde{S}_{\alpha}^{+} = \frac{1}{2}(S_{\alpha}^{z} + 1)S_{\alpha}^{+},
\end{eqnarray}
acting on the state $|-1\rangle_N = \otimes_{\alpha=1}^N |1,-1\rangle_\alpha$ in which a spin$-1$ in the $S_z = -1$ state (i.e. 0 boson occupancy) sits on each site.   Both the gauge phase and spin-flip operators are chosen so that $|\Psi\rangle$ becomes a singlet in the thermodynamic limit \cite{GT,greiter}.  A similar state was studied in \cite{glasser} that, in the large system limit, becomes identical to that proposed in \cite{GT} for the planar geometry. 

When this construction is generalized to the torus the CSL states are again of the form (\ref{wf_droplet_2}), but there is now a three-dimensional space of states corresponding to the three-fold topologically degeneracy of the bosonic Moore-Read states on the torus.  For a rectangular $L_x \times L_y$ system in the Landau gauge this space is spanned by the states \cite{GWW,RR}
\begin{eqnarray}\label{amplitude_torus} 
\Psi_{\alpha}[z_i] &=& {\rm Pf} \left(\frac{\vartheta_{\alpha+1}((z_i - z_j)/L_x|\tau)}{\vartheta_1((z_i - z_j)/L_x|\tau)}\right)\\
&\times&\prod_{i<j}^N \vartheta_1 ((z_i - z_j)/L_x|\tau)
F^{(\alpha)}_{{\rm cm}} (Z)\prod_{i=1}^N  e^{-y_i^2/2},\nonumber 
\end{eqnarray}
where 
\begin{eqnarray}\label{theta}
\vartheta_{\delta}(z|\tau)  
= (-1)^{\tilde{\delta}}\,\sum_{n=-\infty}^{\infty} e^{ [i \pi
\tau (n+a)^2 + 2 \pi i (n+a)(z+b)]}\;,
\end{eqnarray}
are the four Jacobi theta functions where the parameters $(a,b)$ take the values $(1/2,1/2), (1/2,0), (0,0), (0,1/2)$, for $\delta = 1,2,3,4$, respectively, and $\tilde{\delta} = 1$ only for $\delta = 1$, otherwise $\tilde{\delta} = 0$. The parameter $\tau$ is determined by the ratio of the system lengths $\tau = iL_{y}/L_{x}$.   As above, the lattice of points $z = \eta_{nm} = \sqrt{2\pi}(n+im)$ has one flux quantum per plaquette, and, again, when bosons are confined to this lattice the allowed occupancies are 0, 1, and 2. Finally, for even by even lattices the center-of-mass term $F^{(\alpha)}_{{\rm cm}} (Z)$, where $Z = \sum_{i=1}^N z_i$, is taken to be
\begin{eqnarray}\label{center_of_mass}
F_{{\rm cm}}^{(\alpha)} (Z) = \vartheta_{\alpha+1}(Z/L_x|\;\tau),
\end{eqnarray}
to ensure the wave function is periodic for each boson on the lattice with period $L_x$ ($L_y$) in the $x$ ($y$) direction. 

\begin{figure}
\centering
\includegraphics[width=8.5cm]{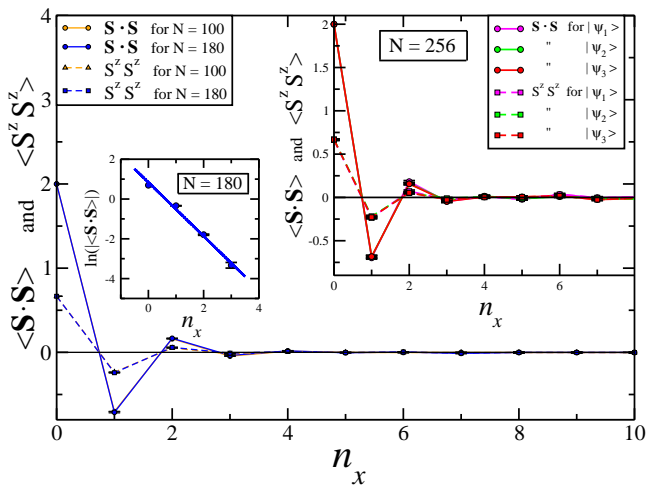}
\caption{\label{corr_droplet_torus}
(Color online)  Spin correlation functions $\langle {\bf S}_{0}\cdot {\bf S}_{0+n_{x}} \rangle$ and $\langle S^{z}_0 S^{z}_{0+n_{x}}\rangle$ versus $n_x$ (lattice spacings in $x$ direction) for droplet CSL states with $N=100$ and $180$ bosons (0 is the droplet center).  The left inset shows a logarithmic plot of $|\langle {\bf S}_{0}\cdot {\bf S}_{0+n_{x}} \rangle|$ with linear fit yielding a correlation length of $\xi = 1.35 \pm 0.14$. The right inset shows the spin correlation functions $\langle {\bf S}_{i}\cdot{\bf S}_{i+n_{x}} \rangle$ and $\langle S^{z}_i S^{z}_{i+n_{x}}\rangle$ for the three states $|\Psi_\alpha\rangle$ on a toroidal lattice of size $16 \times 16$. The spin correlations in these states are indistinguishable within errors.}
\end{figure}

The torus CSL states are again constructed using (\ref{wf_droplet_2}) but with $\Psi$ replaced by one of the three $\Psi_\alpha$ states defined above and with a new gauge factor $G(z_{i})$ with $G(\eta_{n,m}) = (-1)^{(n+m)}$ which takes into account the change from symmetric to Landau gauge.  On the torus, the resulting CSL states $|\Psi_\alpha\rangle$ are exact singlets, even for finite systems \cite{greiter}. This procedure generalizes the Abelian CSL construction on the torus due to Laughlin \cite{laughlin_csl}.  These torus Abelian states were studied by variational Monte Carlo similar to that used here in \cite{nick}.  A general prescription for constructing torus CSL states based on conformal field theory, which includes the non-Abelian case relevant here, was given in \cite{nielsen}. 

We have carried out variational Monte Carlo calculations for both the droplet and torus CSL states.  In all cases the Pfaffian becomes singular when two bosons occupy the same site. However, the wave function remains finite, because the corresponding Jastrow factor ``cancels'' the divergence of the Pfaffian.  In our simulations we treat this singular case by replacing the relevant Jastrow factor and Pfaffian element with 1 for any doubly occupied site, thus correctly reproducing the limiting value of their product. 

{\em Correlations.}---Figure \ref{corr_droplet_torus} shows spin correlation functions $\langle {\bf S}_{0}\cdot {\bf S}_{n_{x}}\rangle$ and $\langle S_{0}^{z} S_{n_{x}}^{z}\rangle$ for the droplet CSL where $0$ is the droplet center and $n_x = x/\sqrt{2\pi}$ is the number of lattice spacings along the $x$ direction.  Results are shown for $N = 100$ and $180$ bosons and it is evident that the correlations for the different system sizes agree.  Note that $\langle S_0^z S_{n_x}^z \rangle \simeq \frac13 \langle {\bf S}_0 \cdot {\bf S}_{n_x}\rangle$ consistent with the approximate singlet nature of the droplet CSL.  We find the absolute value of the spin correlation functions follow a simple exponential law, $|\langle {\bf S}_0 \cdot {\bf S}_{n_x}\rangle| \propto e^{-n_x/\xi}$, even at short distance, consistent with the expectation that the spin-1 CSL can be viewed as a gapped spin liquid.  From our numerics we obtain a spin correlation length of $\xi = 1.35 \pm 0.14$ lattice spacings (see \Fig{corr_droplet_torus} left inset).

\begin{table}[t]
    \begin{tabular}{ | l | r | r | r |}
    \hline
    correlation (exact) & $|\Psi_{1} \rangle$ & $|\Psi_{2} \rangle$ & $|\Psi_{3} \rangle$ \\ \hline
    $\langle {\bf S}_{i}\cdot {\bf S}_{i+\hat x}\rangle$ & $-8/11$ & $-5/4$ & $-16/11$ \\ \hline
    $\langle {\bf S}_{i}\cdot {\bf S}_{i+\hat y}\rangle$ & $-16/11$ & $-5/4$ & $-8/11$ \\ \hline
		    $\langle {\bf S}_{i}\cdot {\bf S}_{i+\hat x + \hat y}\rangle$ & $ 2/11$ & $1/2$ & $2/11$ \\ \hline
    \end{tabular}
\caption{Exact spin correlations for neighboring spins along the $\hat x$ and $\hat y$ directions as well as along the diagonal for $| \Psi_{\alpha} \rangle$, $\alpha = 1,2,3$, on a $2\times 2$ toroidal lattice.}
\label{tab:table1}
\end{table}

Figure \ref{corr_droplet_torus} also shows spin correlation functions for all three CSL states $|\Psi_\alpha\rangle$ on the torus for a $16\times 16$ lattice.  Our results confirm that for a large enough system these correlation functions coincide for all three states within errors (see \Fig{corr_droplet_torus} right inset), and also agree with the droplet correlations.  We note that this is not the case for small system sizes.  For example, for the simple case of a $2 \times 2$ torus all correlation functions can be obtained analytically for all three states with clearly distinguishable results (see Table \ref{tab:table1}). 

\begin{figure}
\centering
\includegraphics[width=8.5cm]{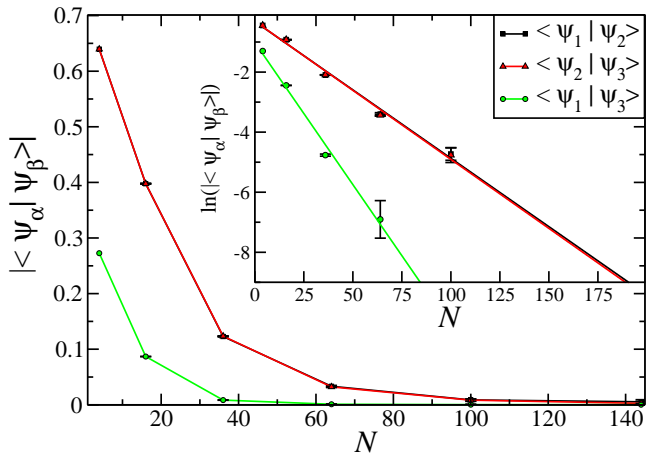}
\caption{ \label{bracket_23_34} (Color online) Normalized overlaps $|\langle \Psi_{1(2)}| \Psi_{2(3)} \rangle|$ and $|\langle \Psi_{1}| \Psi_{3} \rangle|$ for square-shaped systems versus number of lattice sites $N = 4, 16, 36, 64, 100, 144$. The inset shows logarithmic plots of $|\langle \Psi_{1(2)}| \Psi_{2(3)} \rangle|$ and $|\langle \Psi_{1}| \Psi_{3} \rangle|$ versus $N$ with linear fits showing $|\langle \Psi_{1(2)}| \Psi_{2(3)} \rangle|$ becomes exponentially smaller with a decay factor of $\zeta = 0.05 \pm 0.01$, while $|\langle \Psi_{1}| \Psi_{3} \rangle|$ decreases with $\zeta = 0.095 \pm 0.001$. 
 }
\end{figure}

One difference between the droplet and torus CSL states, noted above, is that the droplet only becomes an exact singlet in the thermodynamic limit.  We can see this explicitly by noting that for a singlet state the onsite correlations must satisfy $\langle S_i^z S_i^z\rangle = \frac12\langle S_i^+ S_i^-\rangle = \frac12\langle S_i^- S_i^+\rangle = \frac{1}{3} \langle {\bf S}_i^2\rangle = \frac{2}{3}$.  For the case of a CSL droplet with $4$ bosons we find that, at the droplet center, $\langle S_{0}^{z} S_{0}^{z}\rangle \approx 0.72$ and $\frac{1}{2}\langle S_{0}^{+(-)} S_{0}^{-(+)}\rangle \approx 0.86 (0.42)$.  However, for droplets of 20 bosons or more, all three correlations have nearly converged to the singlet value of $\frac{2}{3}$. In contrast, for the torus our numerics confirm that, even for small system sizes, the expectation values $\langle S_{i}^{z} S_{i}^{z}\rangle$ and $\frac{1}{2}\langle S_{i}^{+(-)} S_{i}^{-(+)}\rangle$ are precisely $\frac{2}{3}$ on all sites.  The fact that the value of these onsite correlation functions provide a nontrivial test of the singlet nature of the spin$-1$ CSL can be contrasted with the spin$-1/2$ case for which $\langle S_i^z S_i^z\rangle$ is always equal to $\frac{1}{4}$.

{\em Orthogonality.}---To establish that the three torus CSL states $|\Psi_\alpha\rangle$ [henceforth assumed normalized] span a three-dimensional space we have calculated their overlap matrix for several square-shaped lattices of sizes $2\times 2, \ldots, 12\times 12$.  In all cases we find the overlap matrix has full rank. Moreover, the off-diagonal matrix elements go to zero exponentially as $e^{-\zeta N}$ where $N$ is the number of lattice sites, with $\zeta = 0.05 \pm 0.01(0.095 \pm 0.001)$ for $|\langle \Psi_1|\Psi_3 \rangle|$  ($|\langle \Psi_1|\Psi_2\rangle|$ and $|\langle \Psi_2|\Psi_3\rangle|$), as shown in Fig.~\ref{bracket_23_34}. Thus, the three states become orthogonal in the thermodynamic limit.   More details are given in the Supplemental Material (SM).

The transformation properties of theta functions under modular transformations imply that, for square-shaped systems, $R_{\pi/2}|\Psi_{1,3} \rangle =| \Psi_{3,1} \rangle$ and $R_{\pi/2}| \Psi_{2} \rangle = |\Psi_2\rangle$, where $R_{\pi/2}$ generates a $\pi/2$-rotation in the plane. We therefore expect $|\langle \Psi_{1}|\Psi_{2} \rangle| = |\langle \Psi_{2}|\Psi_{3} \rangle|$ for any square-shaped system as the numerical results in \Fig{bracket_23_34} confirm.  These symmetry properties are also apparent in the $2 \times 2$ spin correlation functions given in Table \ref{tab:table1}. 

{\em Entanglement entropy.}---The three states $|\Psi_\alpha\rangle$ become orthogonal and possess indistinguishable spin correlations in the thermodynamic limit.  This three-fold topological degeneracy is consistent with the natural hypothesis that the spin-1 CSL state, like the bosonic Moore-Read state on which it is based, is described by $SU(2)_2$ Chern-Simons theory \cite{fradkin1998chern,fradkin}.  To provide further evidence that this is the case we turn to the entanglement entropy.

The Renyi entropy of order $n$ associated with a partitioning of the system into a region $A$ and its compliment $B$ is defined as $S_{n} = -\frac{1}{n-1}\text{ln Tr}(\rho_{A}^{n})\,$, where $\rho_{A} = \text{Tr}_{B}|\Psi \rangle \langle \Psi|$ is the reduced density matrix of region $A$. Ground states of gapped local Hamiltonians exhibit a boundary law scaling which can generically be written in two dimensions for simply-connected regions $A$ as $S_n(\rho_{A}) = \alpha_n L_{A} - \gamma + \cdots $. The leading term is proportional to $L_{A}$, the boundary length of region $A$, while the second term, $-\gamma$, is the topological entanglement entropy (TEE), characteristic of topological phases \cite{LW,KP}.

\begin{figure}
\centering
\includegraphics[width=8.75cm]{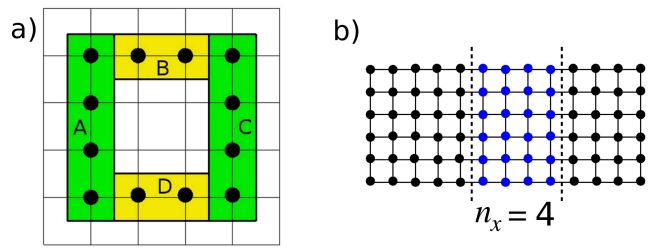}
\caption{ \label{Levin_Wen_cylinder}
(Color online) (a) Example regions $A$,$B$,$C$, and $D$, used in the Levin-Wen construction to isolate the TEE. (b) $12 \times 6$-torus with dashed lines indicating an example region for which the Renyi entropy $S_{2}$ is calculated when partitioning the toroidal system into two cylinders.
 }
\end{figure}

In a topologically ordered state the TEE is determined by the total quantum dimension $\mathcal{D}$, $\gamma = \ln\mathcal{D}$, where $\mathcal{D}$ is defined through the quantum dimension $d_{i}$ of the quasiparticles of the underlying topological field theory: $\mathcal{D} = \sqrt{\sum_{i} d_{i}^2}$. For the spin-1 CSL, based on the continuum bosonic Moore-Read state, we expect the $SU(2)_2$ quantum dimensions of $1, 1, \sqrt{2}$ for which $D = 2$ and $\gamma = \ln 2$.

We proceed by calculating the $n = 2$ Renyi entropy using the replica method \cite{hastings}.  Details are given in the SM.  One way to isolate the TEE is to employ the Levin-Wen \cite{LW} construction  (see \Fig{Levin_Wen_cylinder}(a)), where the area-dependent part cancels from a superposition of four entropies: $-2\gamma = (S_{ABCD} - S_{ADC}) - (S_{ABC} - S_{AC})$.  To combat large error bars, we employed the reweighting scheme of \cite{pei} (see SM). We first choose a relatively small system of size $6 \times 6$ and Levin-Wen regions $A,B,C$ and $D$ as shown in \Fig{Levin_Wen_cylinder}a$)$, resulting in $\gamma = 1.16 \pm 0.08 (1.14 \pm 0.08, 1.04 \pm 0.07)$ for the states $|\Psi_{1(2,3)} \rangle$. The value is above the theoretically expected $\ln 2 \approx 0.69$ but upon increasing the system size to $8 \times 8$, with regions $A(C)$ of size $1\times 6$ and $B(D)$ of size $3\times 2$, we find $\gamma = 0.91 \pm 0.32$ for $|\Psi_{1} \rangle$, consistent with $\gamma$ approaching $\ln 2$ in the thermodynamic limit. 
This is also consistent with the result for $\gamma$ obtained numerically in \cite{glasser} using a bi-cylindrical cut of a CSL state on the cylinder with open boundary conditions. 

\begin{figure}[t]
\centering
\includegraphics[width=8.75cm]{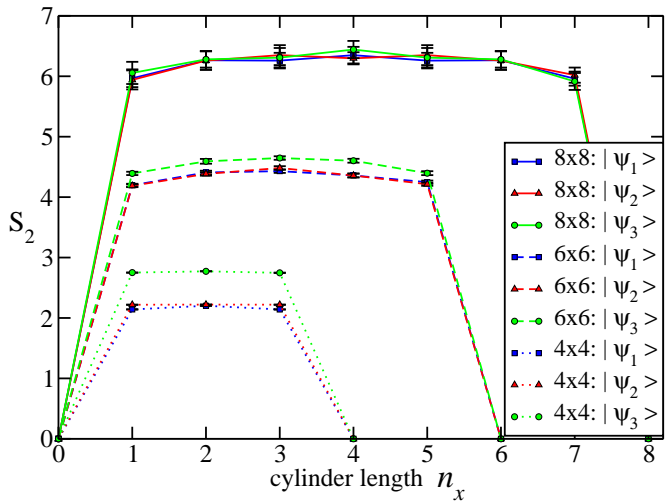}
\caption{ \label{LbyL} (Color online) $S_{2}$ versus the length of the cylindrical region $A$ for square-shaped systems for the three CSL states $|\Psi_\alpha\rangle$.  On a $6 \times 6$ lattice, two wave functions, $|\Psi_{1,2} \rangle$, have identical $S_{2}$ within error bars.  For an $8 \times 8$ lattice, $S_{2}$ is the same for all three ground states within error bars.}
\end{figure}

To identify the modular $\mathcal{S}$-matrix associated with the topological field theory describing the CSL state we follow \cite{zhang_ashvin} and let $|\Xi_i\rangle$ denote the $\hat y$ direction Wilson loop eigenstates associated with quasiparticle of quantum dimension $d_i$ for $i=1,2,3$.  The overlap matrix $V_{ij} = \langle \Xi_{i}|R_{\pi/2}| \Xi_{j}\rangle$ between the (normalized) bases $\{|\Xi_{i}\rangle\}$ and $\{R_{\pi/2}|\Xi_{j}\rangle\}$ (the $\hat x$ direction Wilson loop states) is then related to the modular $\mathcal{S}$ matrix by $V = D^{\dagger} \mathcal{S} D$, where $D$ is a diagonal matrix of phases $D_{jj} = e^{i\Phi_{j}}$ corresponding to the phase freedom of choosing $|\Xi_{j}\rangle$.  It follows that the eigenvalues $R_{\pi/2}$ are the same as those of the modular ${\cal S}$ matrix.

As noted above, for square-shaped systems, $R_{\pi/2}|\Psi_{1,3}\rangle=|\Psi_{3,1}\rangle$ and $R_{\pi/2} |\Psi_2\rangle = |\Psi_2 \rangle$. This, together with the fact that the $|\Psi_\alpha\rangle$ states become orthogonal for large systems, implies the eigenvalues of $\text{R}_{\pi/2}$ are $\{1,1,-1\}$.  The $\mathcal{S}$-matrix for $SU(2)_2$ Chern-Simons theory has the same set of eigenvalues and is the only such rank 3 $\mathcal{S}$-matrix \cite{zwang}.  Thus, if the spin-1 CSL is described by a topological field theory it {\it must} have quasiparticles with quantum dimensions $d_{1,2} = 1$, and $d_3=\sqrt{2}$.

To connect this observation to our numerics, we note that for such a topologically ordered state the TEE becomes state-dependent when $S_{2}$ is calculated on the torus over a (non-simply connected) cylindrical region of length $n_x$ such as that shown in Fig.~\ref{LbyL}(b) \cite{fradkin,zhang_ashvin}, with $S_2 = -\gamma^\prime + \alpha_2 L_A$, where
\begin{eqnarray}\label{gamma}
\gamma^{\prime} = 2\gamma + \text{ln}\bigg(\sum_{j} p_j^2/d_j^2\bigg)\;.
\end{eqnarray}
Here, $p_j = |c_j|^2$ where $| \Psi_{\alpha} \rangle = \sum_{j} c_{j} | \Xi_{j} \rangle$.  We have numerically calculated $S_2$ for all three torus CSL states on square-shaped lattices up to size $8\times 8$.  The results are shown in Fig.~\ref{LbyL}.  We observe first that $S_2$ saturates as $n_x$ increases (for $n_x < \frac12 L_x/(\sqrt{2\pi})$), consistent with these states being possible ground states of a gapped Hamiltonian. Further we find that for large enough systems $S_2$ is the same for all three states $|\Psi_\alpha\rangle$, and thus $\gamma^\prime$ is as well.

The observation that $\gamma^\prime$ is state independent for the $|\Psi_\alpha\rangle$ states, together with the requirement that the eigenvectors with eigenvalue -1 of the known ${\cal S}$-matrix for $SU(2)_2$ \cite{cardy,fradkin,zwang} (for the the phase choice $\Phi_{j} = 0$ for $j=1,2,3$), $(|\Xi_1\rangle-|\Xi_2\rangle)/2 - |\Xi_3\rangle/\sqrt{2}$, and of $R_{\pi/2}$, $(|\Psi_1\rangle - |\Psi_3\rangle)/\sqrt{2}$, must be the same (up to a phase), constrains us to make the identification,
\begin{eqnarray}\label{Psi_ab}
|\Psi_{2,a} \rangle = \frac{1}{\sqrt{2}}(|\Xi_{1} \rangle \pm |\Xi_{2}\rangle),\;\;\;\;|\Psi_{b} \rangle = |\Xi_{3}\rangle \;,
\end{eqnarray}
where $(|\Psi_a\rangle, |\Psi_b\rangle) = (|\Psi_{1} \rangle, |\Psi_{3} \rangle)$ or  $(|\Psi_{3} \rangle, |\Psi_{1} \rangle)$. For both choices it is readily seen that if quasiparticles with $d_{1,2} = 1$ are associated with $|\Xi_{1,2} \rangle$ and the non-Abelian excitation with $d_{3} = \sqrt{2}$ is associated with $|\Xi_{3} \rangle$, (\ref{gamma}) does indeed yield $\gamma^{\prime} = \ln 2$ for {\em all} three states $|\Psi_{\alpha}\rangle$.  Our numerical observation that $\gamma^\prime$ is the same for the states $|\Psi_\alpha\rangle$ is thus consistent with these states being identified as a basis for the three-dimensional topological Hilbert space of an $SU(2)_2$ Chern-Simons theory on the torus.

{\em Conclusion.}---In this work, we investigated several properties of a spin$-1$ CSL on the square lattice. Spin correlations were found to decay exponentially, and, for the torus, become indistinguishable for the states $|\Psi_\alpha\rangle$ for large systems.  We further found these states become orthogonal in the thermodynamic limit.
 
A Levin-Wen construction was used to determine the TEE of the CSL with results consistent with $-\ln 2$ in the thermodynamic limit.  In addition, based purely on symmetry, we argued that the modular ${\cal S}$-matrix of the CSL (if it exists) must be the same as that for the bosonic Moore-Read state.  These observations, together with the observation that for large enough systems the cylindrical entropies for the states $|\Psi_\alpha\rangle$ are all the same, are consistent with the spin$-1$ CSL exhibiting the non-Abelian topological order of $SU(2)_2$ Chern-Simons theory. 

\begin{acknowledgments}
Our MC codes are partially based upon the ALPS libraries\cite{alps1,alps2}. The simulations were run on the SHARCNET clusters.  JW is supported by the National High Magnetic Field Laboratory under NSF Cooperative Agreement No. DMR-0654118 and the State of Florida. 
\end{acknowledgments}

\bibliographystyle{apsrev}                                                 
\bibliography{PAPER_SUPP_dec22_MR_CSL_new3}{} 

\vskip 1cm

{ \bf Supplemental Material: Spin Correlations and Topological Entanglement Entropy in a Non-Abelian Spin-1 Spin Liquid}

{\em 1) Overlaps.}---To check if the three torus CSL states $|\Psi_{\alpha}\rangle$, $\alpha = 1,2,3$, are linearly independent,
we calculate the normalized overlap matrix with entries
$ \frac{\langle \Psi_{\alpha} | \Psi_{\beta} \rangle} 
{||\Psi_{\alpha}|| \; ||\Psi_{\beta}||} $ with $\alpha, \beta = 1,2,3$.
This can be done exactly for small systems, and by variational Monte Carlo calculations for larger systems.

When calculating the overlaps by Monte Carlo we use the following expression,
\begin{eqnarray}\label{eeq3_supp}
\frac{\langle \Psi_{\alpha} | \Psi_{\beta} \rangle}{||\Psi_{\alpha}|| \cdot ||\Psi_{\beta}||} 
&=& 
\frac{\sum_{\{z_{i}\}} \Psi_{\alpha}^{*}(\{z_{i}\}) \Psi_{\beta}(\{z_{i}\})}{\sqrt{\sum_{\{z_{i}\}}|\Psi_{\alpha}(\{z_{i}\})|^2 \cdot 
\sum_{\{z_{i}\}}|\Psi_{\beta}(\{z_{i}\})|^2}}  \nonumber \\
&=& 
\frac{\Lambda_{\alpha\beta}}{\sqrt{\Omega_{\alpha \beta} \cdot \Omega_{\beta \alpha}}},
\end{eqnarray}
where
\begin{eqnarray}\label{eeq4_supp}
\Lambda_{\alpha \beta} = \frac{\sum_{\{z_{i}\}} 
|\Psi_{\alpha}(\{z_{i}\}) \Psi_{\beta}(\{z_{i}\})| \frac{\Psi_{\alpha}^{*}(\{z_{i}\}) \Psi_{\beta}(\{z_{i}\})}{|\Psi_{\alpha}(\{z_{i}\}) \Psi_{\beta}(\{z_{i}\})|}} 
{\sum_{\{z_{i}\}} |\Psi_{\alpha}(\{z_{i}\}) \Psi_{\beta}(\{z_{i}\})|},
\end{eqnarray}
and
\begin{eqnarray}\label{eq34a_supp}
\Omega_{\alpha \beta} = 
\frac{\sum_{\{z_{i}\}} |\Psi_{\alpha}(\{z_{i}\}) \Psi_{\beta}(\{z_{i}\})|  \frac{|\Psi_{\alpha}(\{z_{i}\})|}{|\Psi_{\beta}(\{z_{i}\})|}}
{\sum_{\{z_{i}\}} |\Psi_{\alpha}(\{z_{i}\}) \Psi_{\beta}(\{z_{i}\})|}.
\end{eqnarray}
The quantities $\Lambda_{\alpha\beta}$, $\Omega_{\alpha\beta}$, and $\Omega_{\beta\alpha}$ can be calculated straightforwardly by Monte Carlo sampling over a distribution with weight $|\Psi_\alpha(\{z_i\}) \Psi_\beta(\{z_i\})|$.

{\em 2) Entanglement and replica-trick/SWAP-operator.}---The Renyi entropy of order $n$ associated with a partitioning into complementary regions $A$ and $B$ is defined as
\begin{equation}\label{eq3_supp} 
S_{n} = -\frac{1}{n-1} \text{ln} (\rho_{A}^{n}) \;,  
\end{equation}
where $\rho_{A} = \text{Tr}_{B}|\Psi \rangle \langle \Psi|$ is the reduced density matrix of region $A$. In the limit $n \rightarrow 1$, the von Neumann entropy $S_{1}(\rho_{A}) = -\text{Tr} \rho_{A} \text{ln} \rho_{A}$ is recovered.  For ground states of local Hamiltonians $S_n$ can exhibit a boundary-scaling in region size, which, in two dimensions, can generically
be written as,
\begin{equation}\label{area_supp}
S_{n}(\rho_{A}) = \alpha_{n} L_{A} - \gamma + \ldots\,.
\end{equation}
Here, the leading term is dependent on the ``area'' (or boundary) of region $A$. The subleading term, the topological entanglement entropy (TEE) $-\gamma$, is a signal of topological order and is characterized by the total quantum dimension $D$, defined through the quantum dimensions $d_i$ of the individual quasiparticles of the underlying topological quantum field theory: $D = \sqrt{\sum_{i}d_{i}}$. Note that the TEE $-\gamma$ does not depend on $n$.

Since it is computationally expensive to calculate the von Neumann entropy, we proceed by calculating the second Renyi entropy $S_{2}$. Here, the replica trick approach of \cite{hastings} can be used to efficiently calculate $S_{2}$ by obtaining the expectation value of the so-called $\text{SWAP}$-operator $\langle \text{SWAP}\rangle$ (defined below).  In \cite{hastings} it was shown that $\langle \text{SWAP} \rangle = \text{Tr}(\rho_{A}^{2})$ and consequently
\begin{equation}\label{swap_e_supp}
S_{2} = -\text{ln}(\langle \text{SWAP} \rangle)\;.
\end{equation}

Applying the SWAP-operator to a system with wave function  $|\Psi(\{z_{i}\})\rangle$ requires us to make two independent copies of the system. The total state of the doubled system is then $|\Psi(\{z_{i}\}) \otimes \Psi(\{z_{i}^{\prime}\}) \rangle$. The $\text{SWAP}$-operator now exchanges the degrees of freedom in region $A$ between the two copies, while leaving the the degrees of freedom in the complimentary region $B$ untouched \cite{hastings},
\begin{eqnarray}\label{sw_supp}
\text{SWAP} |\Psi(\{z_{i}\}) \otimes \Psi(\{z_{i}^{\prime}\}) \rangle = |\Psi(\{\tilde{z}_{i}\}) \otimes \Psi(\{\tilde{z}_{i}^{\prime}\}) \rangle\;.
\end{eqnarray}
Here, $\tilde{z}_{i}$ and $\tilde{z}_{i}^{\prime}$ refer to the complex ``particle'' coordinates after the SWAP-operation. For the expectation value $\langle \text{SWAP} \rangle$, we then have,
\begin{eqnarray}
\label{eq11_supp}
\langle \text{SWAP} \rangle = \frac{\sum_{\{z_{i}\}, \{z_{i}^{\prime}\}} |\Psi_{\alpha}(\{z_i\})|^{2} |\Psi_{\alpha}(\{z_{i}^{\prime}\})|^{2} 
\;\times\;
R}
{\sum_{\{z_{i}\}, \{z_{i}^{\prime}\}} |\Psi_{\alpha}(\{z_i\})|^{2} |\Psi_{\alpha}(\{z_{i}^{\prime}\})|^{2} }
\end{eqnarray}
with the measurement/estimator
\begin{eqnarray}\label{eq11a_supp}
R = \frac{\Psi_{\alpha}(\{\tilde{z}_{i}\}) \Psi_{\alpha}(\{\tilde{z}_{i}^{\prime}\})}{\Psi_{\alpha}(\{z_{i}\}) \Psi_{\alpha}(\{z_{i}^{\prime}\})}\;.
\end{eqnarray}
It is clear from  \eqref{eq11a_supp} that the measurement depends on the state of the two systems after the swapping of the degrees of freedom.

Since the number of bosonic configurations grows exponentially with the size of the subsystem $A$, we have exponentially large fluctuations in the measurement and, as implied by the area law, only an exponentially small part of the original bosonic configurations will lead to a non-zero measurement of $R$. Most measurements will be zero.

Thus, in order to combat large errors, we employ the re-weighting scheme of Pei {\it et al.} \cite{pei}.  This re-weighting scheme splits the expectation value $\langle \text{SWAP} \rangle$ into two parts,
\begin{eqnarray}\label{eq11b_supp} 
\langle \text{SWAP} \rangle = \langle \text{SWAP}_{sign} \rangle \times \prod_{j}^{m}\langle \text{SWAP}_{amp} \rangle_{j} \;. 
\end{eqnarray}
The first part is the sign-dependent part of the SWAP-operator, the second part is itself a product over all contributions to the amplitude of SWAP.

We find for the sign-dependent part
\begin{eqnarray}\label{eq12_supp} 
\langle \text{SWAP}_{sign} \rangle   
&=& \frac{\sum_{\{z_{i}\}, \{z_{i}^{\prime}\}} 
|A(\tilde{z}_{i}, \tilde{z}_{i}^{\prime}, z_{i}, z_{i}^{\prime})|  
e^{i\Phi(\{z_{i}\}, \{z_{i}^{\prime}\})}} 
{\sum_{\{z_{i}\}, \{z_{i}^{\prime}\}} |A(\tilde{z}_{i}, \tilde{z}_{i}^{\prime}, z_{i}, z_{i}^{\prime})|,
} \nonumber \\ 
&& 
\end{eqnarray}
where we defined the weight
\begin{eqnarray}\label{eq12a_supp} 
|A(\tilde{z}_{i}, \tilde{z}_{i}^{\prime}, z_{i}, z_{i}^{\prime})| &=&   
|\Psi_{\alpha}(\{\tilde{z}_{i}\}) \Psi_{\alpha}(\{\tilde{z}_{i}^{\prime}\}) \Psi_{\alpha}(\{z_{i}\}) \Psi_{\alpha}(\{z^{\prime}_{i}\})| \nonumber \\ 
&& 
\end{eqnarray}
and $\Phi(\{z_{i}\}, \{z_{i}^{\prime}\})$ is the phase of $R$.

The amplitude-dependent part itself is a product. This allows us to express the quantity $\langle \text{SWAP}_{amp} \rangle$ to be evaluated as the product of a series of ratios so that the evaluation of each ratio only suffers from a much smaller fluctuation.  This can practically be done by introducing $r_{j} \in [0,1]$ as a series of powers satisfying $r_{j} < r_{j+1}$, $r_{1} = 0$ and $r_{m+1} = 1$.

Defining the weight
\begin{eqnarray}\label{eq13_supp} 
\tilde{W}_{j}(\{z_{i}\}, \{z_{i}^{\prime}\}) &=& 
|A(\tilde{z}_{i}, \tilde{z}_{i}^{\prime}, z_{i}, z_{i}^{\prime})|^{1-r_{j+1}} \times \nonumber \\ 
&\times& 
\Big(|\Psi_{\alpha}(\{z_i\})|^{2} |\Psi_{\alpha}(\{z_{i}^{\prime}\})|^{2}\Big)^{r_{j+1}}\;, 
\end{eqnarray}
we have
\begin{eqnarray}\label{eq14_supp} 
\langle \text{SWAP}_{amp} \rangle &=& \prod_{j}^{m} \frac{\sum_{\{z_{i}\}, \{z_{i}^{\prime}\}} \tilde{W}_{j}(\{z_{i}\}, \{z_{i}^{\prime}\}) 
\times 
|R|^{r_{j+1} - r_{j}}} 
{\sum_{\{z_{i}\}, \{z_{i}^{\prime}\}} \tilde{W}_{j}(\{z_{i}\}, \{z_{i}^{\prime}\})}\;. \nonumber \\ 
&& 
\end{eqnarray}

If $r_{j+1} - r_{j}$ are chosen to be sufficiently small, each term in the above product can be evaluated easily and will {\it not} suffer from large fluctuations. It is easy to see that the crucial feature of the re-weighting scheme is that now the weights contain the amplitudes of the wave functions {\it after} the swapping of the degrees of freedom.  Consequently, all measurements are now non-zero.

We remark that if we calculate the entanglement entropy  $S_{2}$ over a cylindrical area, using \eqref{eq11_supp} is sufficient, whereas the re-weighting scheme is necessary for the Levin-Wen construction in order to obtain sufficiently small error bars.

For all Levin-Wen calculations in this Letter, a step width of $r_{j+1} - r_{j} = 0.2$ was used. Thus the amplitude-dependent part of the SWAP-operator \eqref{eq11b_supp} is a product consisting of $5$ factors.  Adding a single calculation for the sign-dependent part of SWAP, we find that each of the four Levin-Wen regions requires six different calculations, the product of six ``partial'' expectation values corresponds to the ``full'' expectation value $\langle \text{SWAP} \rangle$ for the respective area.

For the bosonic spin-$1$ Moore-Read spin liquid, we observe that the error bars for all expectation values of the amplitude-part of the SWAP-operator converge relatively fast and are of order $10^{-4(5)}$.  However, the error stemming from the sign part $\langle \text{SWAP}_{sign}\rangle$ is significantly larger, and therefore mainly responsible for the (total) errors given in the Letter.

\end{document}